\documentstyle[12pt]{article}
\begin{document}
\def\bra{\langle}
\def\ket{\rangle}
\def\cA{{\cal A}}
\def\cE{{\cal E}}
\def\cH{{\cal H}}             
\def\cU{{\cal U}}
\def\cH{{\cal H}}
\def\cQ{{\cal Q}}
\def\cZ{{\cal Z}}
\def\Qp{{\cQ_p}}
\def\Zp{{\cZ_p}}
\def\N{{I\kern-.25em{N}}}
\def\R{{I\kern-.25em{R}}}
\def\Z{{Z\kern-.5em{Z}}}
\def\V{{V\kern-.7em{V}}}
\def\ie{i.e.\,}
\def\be{\begin{equation}}
\def\ee{\end{equation}}
\def\d{\partial}
\def\bra{\langle}
\def\ket{\rangle}
\def\PRL{{\em Phys. Rev. Lett.}}
\def\PR{{\em Phys. Rev.}}
\def\lr{{\rm L}^2({\rm R})}
\def\Ex{{\rm M}}              
\def\Cov{{\rm R}}             
\title{Scale-dependent functions in statistical hydrodynamics:
a functional analysis point of view}
\author{M.V.Altaisky\\ Joint Institute for Nuclear Research, 
Dubna, 141980, Russia \and Space Research Institute, Profsoyuznaya 
84/32, Moscow, \\ 117810, Russia}
\date{Revised: Aug 3, 1997}
\maketitle
\begin{abstract}
Most of dynamic systems which exhibit chaotic behavior are also known
to posses self-similarity and manifest strong fluctuations of all possible
scales.The meaning of this terms is not
always the same. In the present note we make an attempt to formulate the
problem in the framework of functional analysis. The statistical hydrodynamics
is taken as a vivid physical example.
\end{abstract}
\section{Introduction}
The problem of adequate mathematical description of hydrodynamic turbulence
is one of the oldest but not yet solved problems in physics. The transition 
from laminar fluid flow to highly irregular chaotic regime was first 
discussed yet by Leonardo da Vinci. Later, when the Navier-Stokes 
equation (NSE)
\be
\d_t {\bf v} + {\bf v}\cdot\nabla{\bf v}=-\nabla p +\nu\Delta {\bf v}, 
\quad \nabla\cdot{\bf v}=0,
\label{nse}
\ee
which describes an incompressible fluid flow was written down, it was 
believed that (\ref{nse}) itself may contain all turbulence. 
It seems quite natural in the light of modern developments in dynamical 
chaos, that even systems with a few degrees of freedom sometimes show 
unpredictable chaotic behavior. 

The turbulence problem is much more complicated. The velocity 
{\em field} ${\bf v}({\bf x},t)$ has $3\times 3D\ continuum$ degrees of 
freedom -- thus we have a field theory problem. The dynamical object  
-- the velocity field ${\bf v}({\bf x},t)$ itself is a square-integrable 
function defined on $\R^3\times\R^1$ space.

The crucial step in turbulence theory was done by A.~N.~Kolmogorov 
in a number of short papers \cite{K41a,K41b,K41c}, known as K41 theory. 
It was argued, that the turbulence, as a chaotic phenomenon, should be 
described in terms of random functions ${\bf v}({\bf x},t,\cdot)$ 
(The ideas of statistical description of turbulence have been already 
discussed by Taylor\cite{Tay35}). {\em It was suggested to consider the 
turbulence as a multi-scale phenomenon, formed of velocity fluctuations 
of all possible scales.} The typical size $L$ of dominating fluctuations 
of average amplitude $U$ is related to the Reynolds number 
$Re = \frac{UL}{\nu}$, which determines the transition from 
laminar to turbulent flow. Following \cite{Frish95} we present 
K41 hypothesis in the form 
\begin{description}
\item[H1] In the limit $Re \to \infty$, all possible symmetries of 
the NSE, usually broken by the mechanisms reducing the turbulent flow, 
are restored in a statistical sense at small scales and away from 
boundaries.
\item[H2] Under the same assumptions as in H1, the turbulent flow is 
self-similar at small scales, i.e.:
\be
\delta v (r,\lambda l) \stackrel{law}{=} \lambda^h \delta v (r,l)
\label{k41}
\ee
where $v (r,\lambda l)  = v (r+\lambda l)- v (r)$. (The equality-in-law 
means the coincidence of all statistical momenta.)
\item[H3] Under the same assumptions as in H1 and H2, the turbulent 
flow has a finite non-vanishing rate $\epsilon$ of dissipation of energy 
per unit of mass.
\end{description}
Using these three assumptions Kolmogorov has derived the two-thirds 
law
\be
\bra (\delta v(l))^2\ket = C \epsilon^{2/3}l^{2/3},\label{l23}
\ee
the basic empirical law of fully developed turbulence.

The scaling behavior (\ref{l23}) of velocity field $v$ 
means that we have to deal with functions of typical behavior 
\be
|v(r+l)-v(r)| \sim l^h 
\label{law}
\ee
with $h\le1$, {\em i.e. with non-differentiable functions}; 
with nontrivial H\"{o}lder exponent $h=1/3$, if H3 hypothesis is 
strictly valid. 
The H\"{o}lder condition $|v(r+l)-v(r)| \le l^h$ is not very rare in 
physical problems: it can be found in condensed matter physics, quantum 
field theory etc. --- it becomes very significant only since the 
resolution ($l$) itself becomes a physical parameter, and that is why we 
ought to consider some objects of the form $v(x,l,\cdot)$, which are not 
yet defined. 
This problem is actively discussed amongst the turbulence community, but up 
to the authors knowledge is beyond functional analysis research, except 
for the problems directly related to wavelet analysis \cite{Holsh94}. 

In present paper we investigate the problem of resolution dependent 
functions. The research was highly inspired by the fact, that there are 
at least two approaches to resolution dependent objects. First, 
based on the decomposition with respect to the group of affine 
transformations, is known as wavelet analysis. Second approach was 
exported from quantum field theory to turbulence and is based 
on the consideration of the Fourier transforms of functions 
cut at some large momentum $k=\Lambda_{cut-off}$. 

An attempt 
to consider the problem as a whole and join these two approaches 
is presented below.
\section{Problem}
Self-similarity is a synonym of scale-invariance. To be scale-invariant
means to have same properties at different scales. Classical fractals
are scale-invariant by construction. 
Brownian motion
is self-similar: if we look at the trajectory of Brownian particle at
different resolution of a microscope we will observe more or less the
same picture.

As physical systems are considered, the word self-similarity is more
frequently attributed to their dynamics than geometry.

The self-similarity of hydrodynamic velocity field fluctuations
$\bra (\delta v(l))^2 \ket \sim l^{2/3}$ is attributed to behavior of
turbulent velocity field measured at different spatial scales. For
hydrodynamical velocity field it is physically clear, that the
measurement at scale $l_0$ necessarily implies averaging of molecular
velocities over certain space domain of typical size $l_0$. This
procedure can be generalized to ``an averaging of a function up to
scale l'' \cite{DG96}
\be
\phi_l({\bf x}) = l^{-D} \int_{|y|<l}\phi({\bf x}-{\bf y})d^D {\bf y}.
\label{iov}
\ee
There are at least two conjectures here:
\begin{enumerate}
\item The existence of ``true'' (with no scale) field
$\phi_l({\bf x}): l \to 0$.
\item The homogeneity of the measure $d\mu({\bf y})=d^D {\bf y}$.
\end{enumerate}
Physically, it is quite clear, that two different fields
$\phi_l({\bf x})$ and $\phi_{l'}({\bf x})$ live in different functional
spaces if $l\ne l'$. It is meaningless, say, to subtract their values.
Therefore, the velocity field of hydrodynamic turbulence is something
more than a random vector field defined on $R^D\times R$.

To characterize the turbulent velocity at certain point ${\bf x}$ we ought
to know the collection of velocity values $\{ \phi_l({\bf x}) \}$ at a set 
of scales labeled by $l$. The set of scales may be countable
$$ l = l_0, kl_0,k^2l_0, k^3l_0, \ldots, $$
say $k=1/2$ for period-doubling decomposition, or continuous.

To characterize this set it was proposed in \cite{DG96} to use a collection
of unit fields at different scales -- a ``reference field''
$\{ {\cal R}_l(x)\}$. The principal question arising here, is how to describe
the interaction of the fluctuations of different scales. Practically, this
problem is often coup with by decomposition of ``real'' (out-of- scale)
field into slow (large-scale) and fast (small-scale) components.
$$ \phi = V + v,\quad \hbox{where} \quad \Ex v  = 0$$
($\Ex $ means the averaging, or mathematical expectation, here and after.)
In this approach the slow component $V$ governs the equation for $v$ and
the even-order moments of $v$ contribute to the equation for $V$.

From the other hand, as we know from both Kolmogorov's theory and RG
approach, there are no absolute scales in hydrodynamics, except for
dissipative scale and external scale (the size of the system). So, at
least at this middle -- the Kolmogorov's range -- the equations should
be scale-covariant.
The structure which reveals here looks like a fiber bundle over
$R^D$, with leaves labeled by scale. The fluctuations of different scales
may be dependent or independent for various physical situations; but
at least some similarity should be present.

To construct a basic system on this bundle let us follow the ideas of
{\em multi-resolution analysis} \cite{Ma86}.
Let us construct a system of functional subspaces $\{V_i:V_i\subset \cH\}$,
Where $\cH$ is a space of physical observables.
Let the system $\{ V_i\}$ be such, that
\begin{enumerate}
\item $$\ldots \subset V_2 \subset V_1 \subset V_0 \subset \ldots $$
      \label{MRA1}
\item $$ \cap_\infty V_i = \emptyset,\quad \overline{\cup_\infty V_i} = \cH$$
\item Subspaces $V_i$ and $V_{i+1}$ are similar in some way.
      $$f(x)\in V_j \leftrightarrow f(kx)\in V_{j-1},$$ if
      $\{ \phi_i(x) \}_{i\in I}$ forms basis in $V_j$ then
       $\{ \phi_i(kx) \}_{i\in I}$ forms basis in $V_{j-1}$
\end{enumerate}
If the sequence $\{ V_i\}$ is bounded from above, the maximal subspace
is called the highest resolution space. Let it be $V_0$. Then any
function from $V_1$ can be represented as a linear span of $V_0$ basic
vectors. Therefore, the basis $\phi_0$ of the highest resolution space
provides a basis for a whole bundle.

It seems attractive to generalize MRA axioms to the case of continuous set
of scales. Since the chain of subspaces described above implies sequential
coarse graining of the finest resolution field, some details are being
lost in course of this process. The lost details can be stored into the
set of orthogonal complements
\be
V_0 = V_1 \oplus W_1,\quad V_1 = V_2 \oplus W_2, \ldots \label{MRA2}
\ee
So, $\cH = \overline{\sum_k W_k}, W_k\cap W_j = \emptyset$ if $j\ne k$,
and the system  $\{ W_i\}$
can be considered instead of  $\{ V_i\}$. The former has the structure
of $\sigma$-algebra, and thus it is suitable integration.

The fact, that velocities $\phi_l$ and  $\phi_{l'}$ live on different leaves
suggest that their Fourier decomposition should be taken separately
at their leaves
\be
\phi_l(x) = \int \exp(
                -\imath k^{(l)}x^{(l)}
                     ) \tilde\phi_l(k^{(l)})
   d\mu_L^{(l)}(k^{(l)}), \label{sft}
\ee
or some other care should be taken about it in order not to mix 
fluctuations with the same wave vectors but contributing to different scales.
The choice of the left-invariant measure $d\mu_L^{(l)} (k^{(l)})$ is restricted
by the fact, that velocity components measured at certain scale are
mainly concentrated close to this scale. So the measure can
be expressed as
$ d\mu_L^{(l)}(k) = dk W(|l^{-1}-ak|),$ where $W$ vanish at $x\to\pm\infty$,
$a$ is a constant.

The decomposition (\ref{sft}) turns to be a kind of Gabor
transformation \cite{Gabor}.
The measure can $d\mu$ can be explicitly scale-dependent, since the
probability spaces $(\Omega_l, \cU_l,P_l)$ depend on scale..

At this point we arrive to the difference from standard wavelet
approach, where the probability space is completely determined at
finest resolution scale. However, if we accept the hypothesis that
fluctuations of different sizes are statistically independent, we have
to define the probability spaces separately.

\section{Wavelet realization}
We start our construction of multi-scale description with
simplistic one-dimensional case, which is however of practical
importance since only one component of velocity field is often
measured.

Any square integrable function $f(t)\in\lr$ can be represented as a
decomposition with respect to the representations of affine group
\be
t' = at+b
\label{ag}
\ee
\be
f(t) = C_\psi^{-1}\int \frac{1}{\sqrt a}\psi\left(\frac{t-b}{a}\right)
W_\psi(a,b)\frac{dadb}{a^2},
\label{wl}
\ee
which is just another form of the partition of unity with respect
to representation of affine group acting on a Hilbert space \cite{GMP}
$$
\hat 1 = C_\psi^{-1}\int_G U^*(g)|\psi\ket d\mu_L(g)\bra\psi| U(g),
$$
which holds if there exists such $\psi \in \cH$, that
$$
C_\psi := \frac{1}{\|\psi \|_2^2} \int_G |\bra\psi,|U(g)\psi\ket|^2 d\mu_L(g)
<\infty 
$$
holds; $ d\mu_L(g)$ denotes left-invariant measure on $G$.
The scalar products
$$ W_\psi(g)f := \bra f, U(g)\psi \ket$$ are known as wavelet coefficients.

For the case of affine transformation group (\ref{ag}) the normalization
constant $C_\psi$ can be easily evaluated in Fourier space
\be
C_\psi = \int_{-\infty}^{\infty} \frac{|\tilde\psi(k)|}{k}dk =
       2\int_{0}^{\infty} \frac{|\tilde\psi(k)|}{k}dk,
\ee
where $\psi(t) = \frac{1}{2\pi}\int \exp(\imath k t) \tilde\psi(k)dk.$
For the affine group (\ref{ag})
\be
U(a,b)\psi(x) := \frac{1}{\sqrt a}\psi\left(\frac{x-b}{a}\right);
\quad d\mu_L(a,b) = \frac{dadb}{a^2}
\label{rep}
\ee
The corresponding wavelet coefficients are
\be
W_\psi(a,b) = \int \frac{1}{\sqrt a}
\overline{\psi\left(\frac{t-b}{a}\right)}f(t)dt
\label{wt}
\ee
For practical analytical calculations it is often more efficient
to calculate wavelet coefficients in Fourier representation, since
multiplication should be done then instead of convolution:
\be
W_\psi(a,b)f = \frac{1}{2\pi}\int \sqrt{a} \exp(\imath k b)
              \overline{\tilde\psi(ak)}\tilde f(k)dk
\ee
and similar for reconstruction (\ref{wl}).
The decomposition (\ref{wt}) and its inverse (\ref{wl}) are known as
wavelet analysis (See e.g. \cite{Daub} for general review.)
The scalar product (\ref{wt}) is readily seen to be the projection
of the original ``no-scale'' function $f$ to the subspace $W_a$ of MRA
system \ref{MRA2}.

If $f$ is a random function defined on a probability
space $(\Omega, \cA, P)$, the wavelet coefficients
\be
W_\psi(a,b,\cdot) = \int \frac{1}{\sqrt a}
\overline{\psi\left(\frac{t-b}{a}\right)}f(t,\cdot)dt
\label{wtr}
\ee
are also random; the stochastic integration is implied.
As it is known from the theory of stochastic processes, any
random function $\xi(t,\omega), t\in\R^1, \omega\in\Omega$ can
be represented in a spectral form
\be
 \xi(t) = \int \phi(t,\lambda)\eta(d\lambda),
\label{sd}
\ee
where $\phi(t,\lambda)$ is a square-integrable function, $\eta(d\lambda)$
is a stochastic measure
$$
\Ex{\eta(d\lambda)} = 0,\quad
\Ex{|\eta(d\lambda)|^2} = F(d\lambda).
$$
The particular form of the spectral representation (\ref{sd}) is
Fourier representation
$$ \xi(t) = \int \exp(\imath\lambda t) \eta(d\lambda).$$
In a multi-scale case we can introduce a collection of random
processes, each of which belongs to its own leaf of MRA, labeled by
resolution parameter $a$
$$\xi_a(t) = \int \phi_a(t,\lambda)\eta_a(d\lambda).$$
The peculiarity of stochastic case is, that in contrast to the decomposition
of a function with respect to given basic wavelet $\psi(t)$, the function
$\phi_a(t,\lambda)$, which depends on both the properties of random
process itself and filtering properties of measuring equipment, is not
known exactly. Therefore, we have to construct a decomposition, which
has a well defined limit to deterministic case and can be tackled without
the exact specification of the form of basic wavelet $\psi$.

A straightforward way to do it, is to factorize the scaling part of
the left-invariant measure from ``purely stochastic part''
$$
\xi(t) = \int \phi_a(t,\lambda)\eta_a(d\lambda)
\frac{da}{a}
$$
or in the spectral form
\be
\xi(t) = \frac{1}{2\pi C_\psi}\int e^{\imath\lambda t}\tilde\psi(\lambda a)
\eta_a(d\lambda)\frac{da}{a},
\label{gsp}
\ee
where $\eta_a(d\lambda)$ can be considered as generalized
wavelet coefficients, the existence of which does not require the
existence of ``no-scale'' prototype.The left invariant measure
$d\mu(a) = da/(2\pi a)$ on the multiplicative group $x'=ax$ instead
of (\ref{ag}), since translations are already incorporated into
the exponent; $2\pi$ multiplier is introduced for the convenience of
Fourier transform.

The representation (\ref{gsp}) was constructed only to meet the
non-stochastic limit and is not unique. For instance, we can redefine the
spectral measure to incorporate both the properties of the signal and
that of measuring apparatus
$$\xi(t) = \int e^{\imath\lambda t}\hat\eta_a(d\lambda)\frac{da}{a}.$$
The specific energy-per-scale density can be easily evaluated then
\be
\begin{array}{lcl}
\int E(a)da &=& \Ex \int \xi(t)\bar\xi(t)dt \\
            &=& \Ex \int \exp(\imath t (\lambda_1-\lambda_2))
\eta_{a_1}(d\lambda_1)\bar\eta_{a_2}(d\lambda_2)d\mu(a_1)d\mu(a_2)dt \\
            &=& \Ex \int\eta_{a_1}(d\lambda)\bar\eta_{a_2}(d\lambda) d\mu(a)\\
            &=& \int F(a_1,a_2;d\lambda) d\mu(a_1) d\mu(a_2) = E
\end{array}
\label{E}
\ee
The equation (\ref{E}) is a stochastic counterpart of a well known
equation for wavelet energy per scale
\be
E(a) = C_\psi^{-1}\int \frac{|W(a,b)|^2}{a^2}db =
     \frac{1}{2\pi C_\psi}\int\frac{|\hat W(a,k)|^2}{a}dk.
\label{Ek}
\ee
Similar spectral characteristics have been already used for the 
analysis of turbulent data \cite{Ast96}.

If random functions are considered the equation (\ref{Ek}) can be rewritten
in RG-like form
\be
\frac{\d E}{\d\ln a} = \frac{1}{2\pi C_\psi}  \Ex \int |\hat W(a,k)|^2dk,
\label{rg1}
\ee
where $\hat W(a,k)=\overline{\tilde\psi(ak)}\tilde f(k)$ can be understood
as the original noisy signal $f$ perceived by filter $\psi$, i.e. as generalized
wavelet coefficients,
not necessary having ``no-scale'' prototypes, cf.eq.\ref{gsp}.

The logarithmic derivative at the l.~h.~s. of (\ref{rg1}) is exactly 
that of renormalization group equation. However, the r.~h.~s 
of this equation was obtained without any cutoff assumptions, it was 
formally derived from the decomposition of initial (``infinite resolution'') 
signal filtered by window function $\psi$. 

Physically, the existence of the infinite resolution limit is often 
meaningless. For instance, the behavior of electromagnetic coupling 
constant at Plank scale is just a nonsense. The same happens in 
hydrodynamics when the scale parameter comes close to the mean free 
path. However, there is a principal difference between hydrodynamical 
turbulence description and quantum field theory. The scale $l$ 
(resolution) becomes a physical measurable parameter in hydrodynamics, 
and thus $\hat W(a,k)$ can be considered as Fourier components of 
the velocity field fluctuations of different typical sizes. 
The study of the behavior of $\hat W(a,k)$ can provide more 
consistent picture of what happens at different scales than 
standard Fourier decomposition.

\section{High frequency cut-off}
Being utmost scale-invariant at moderate scales, the behavior of
turbulent velocity field changes when approaching the smallest
and largest scales, between which the hydrodynamical description is
valid. The former is the Kolmogorov dissipative scale ($\eta$), the
latter is the size of the system. The size of the system can often be
set to infinity with no harm to physics; whilst the dissipative scale is
of physical importance, since the energy dissipation rate $\bar\epsilon$
is very constant which determines the turbulence behavior in inertial
range.

That is why in RG, as well as in spectral calculations, the cut-off
dependent velocity field is often considered
\be
v^{<}_F(x) = \frac{1}{(2\pi)^d}\int_{|k|<F} \exp(\imath k x) \tilde v(k) dk.
\label{vcut}
\ee
The cumulative energy of all harmonics with wave vectors less or
equal to the cut-off value $F$ is one of the main spectral characteristics
of developed turbulence
\be
\cE(F) = {1\over2} \Ex \int \overline{v^<_F(x)}v^<_F(x)d^dx =
              {1\over2} \Ex \int_{|k|<F} \overline{\tilde v(k)}\tilde v(k)
              \frac{d^dk}{(2\pi)^d}.
\label{ce}
\ee
Similarly, we can consider the cumulative energy of all velocity
fluctuations with typical size greater or equal to a given $A$.
For simplicity let us consider a one-component velocity field
considered as a function of time
\be
\begin{array}{lcl}
E(A) &=& {1\over2} \Ex \int_{|a|\le A} \overline{v(t)}v(t) dt
      =  {1\over C_\psi} \Ex \int_{a=A}^\infty |W_\psi(a,b)v|^2
         \frac{dadb}{a^2} \\
     &=& C_\psi^{-1}\int_{A}^\infty \frac{|\tilde\psi(y)|^2}{y}dy
         \cdot \Ex \int |\tilde v(k)|^2 \frac{dk}{2\pi}
\end{array}
\label{Ea}
\ee
where $$\lim_{A\to0} 2\int_{a=A}^\infty \frac{|\tilde\psi(y)|^2}{y}dy = C_\psi$$and $$ E ={1\over2}  \Ex \int |\tilde v(k)|^2 \frac{dk}{2\pi} $$ is
the total energy of all velocity fluctuations.

For non-vanishing $A$
\be E(A) = F(A) E,\quad  \hbox{where}, \quad 
F(A) = \frac{
            \int_{A}^\infty \frac{|\tilde\psi(y)|^2}{y}dy
            }
            {\int_{0}^\infty \frac{|\tilde\psi(y)|^2}{y}dy}.
\ee

For a better definiteness, let us calculate the filtering function $F(A)$
for a particular family of vanishing momenta wavelets 
\be
\psi_n(x) = (-1)^n \frac{d^n}{dx^n} \exp(-x^2/2), \quad
\tilde\psi_n(k) = \sqrt{2\pi}(-\imath k)^n \exp(-k^2/2),
\label{vmf}
\ee
often used for studying of hydrodynamical velocity field\cite{ABM91,ABM93}.

The normalization constant for this family is
$$C_n = 2\pi \int_{-\infty}^{\infty} k^{2n-1}e^{-k^2}dx = 2\pi\Gamma(n)$$
and so
\be
F_n(A) = \frac{\int_{A^2}^\infty y^{n-1} e^{-y} dy}{\Gamma(n)}.
\ee
The derivative of cumulative energy with respect to logarithmic
measure $da/a$ is
$$
\frac{\d E}{\d \ln A} = E\frac{\d F_n(E)}{\d \ln A}
        = - \frac{\d A^2}{\d \ln A} f_n(A^2)
$$
where $f_n(x) = x^{n-1}e^{-x}/\Gamma(n)$.
So we arrive at RG like equation
\be
\frac{\d E}{\d \ln A} = - \frac{2A^{2n}\exp(-A^2)}{\Gamma(n)} E
\ee
For sufficiently small $A$ the exponential term is close to
unity, and thus the behavior is approximately proportional
to $A^2$.
\section{Self-similarity in a bounded domain of scales and Scale-relativity}
A direct approach to scale dependent functions, which was proposed by 
Nottale\cite{Not}, is based on the assumption that physics is 
scale dependent, but scale covariant. The former means to give up 
the differentiability and consider functions $v(x,l)$ with  
$\displaystyle \lim_{l\to0}\frac{\d v(x,l)}{\d x}$ not necessary existing;
the latter imposes generalized scale covariant equation 
\be
\frac{\d v(x,l)}{\d \ln l} = \beta (v(x,l)), \label{b}
\ee 
which states that scale behavior of a scale-dependent function is 
completely determined by the value of this function at a given 
scale.

The benefit of scale relativity\cite{Not} approach is to provide 
a possibility to account for the processes which admit self-similarity 
only at a limited domain of scales
$$ v(\lambda l) = \lambda^h v(l), \eta \ll l \ll \Lambda.$$
For a homogeneous scale dependent function (mono-fractal) 
$v = v_0(\lambda/r)^\delta$ where $\delta$ is independent of scale $r$. 
To get the additive, rather than multiplicative form for two sequential 
scale transformations, it is convenient to express them in logarithmic 
form : 
\be
\ln \frac{v(r')}{v_0} = \ln \frac{v(r)}{v_0} + \V \delta(r), 
\label{log}
\ee
where $\delta(r') = \delta(r) = \delta, \V = \ln(r/r')$. 
In real hydrodynamic turbulence the multi-fractal behavior is observed 
$$ \delta = \delta(r).$$
In the inertial range $ \eta \ll r \ll \Lambda$ the exponent $\delta$ 
is practically a constant, but closer to the limiting scales 
$\eta$ and $\Lambda$ the dependence of scale becomes significant.

The logarithmic form (\ref{log}) suggests a direct generalization 
on the multi-fractal case $\delta = \delta(r)$, which is like 
the generalization of Galilelian transform to Lorenzian transform:
\be
\begin{array}{lcl}
X' &=& \Gamma(\V) [ X - \V T] \\
T' &=& \Gamma(\V) [ A(\V) X + B(\V) T],
\end{array} \label{sr}
\ee
where $T = \ln (l/l_0), X(T) = \ln \Ex (v_l/v_0)$, see \cite{DG96} 
for details. 

The composition of two scale transformations of this type behaves 
like a composition of two Lorentzian boosts: when approaching 
the unpassible limit, then instead of ``Galilelian'' 
\be
\delta(r) = \delta_0\bigl(1-\ln^2(\Lambda/r)/ \ln^2(\Lambda/\eta)
            \bigr)^{-1/2}
\ee
for the power behavior of velocity field.
The transformation group (\ref{sr}) can be used to construct a 
wavelet decomposition at a limited domain of scales. This, however, 
is the subject of the next paper (in preparation).

\section{Conclusion}
In present paper we give a mathematical framework for the analysis of 
functions which depend on scale. Usually, the scale-dependent functions 
express the value of a certain physical quantity measured at a point $x$ 
by averaging over a box of size $l$ centered at $x$. Such functions are 
often used in hydrodynamics, geophysics, signal analysis. One of the 
most known ways of treating resolution-dependent functions is to identify
the size of the box ($l$) with the inverse wave number of Fourier transform 
$k^{-1}=l$. The higher wave numbers are then cutted off. Sometimes this 
procedure leads to confusion (many problems of field theory approach 
to hydrodynamic turbulence originate from this confusion).

In our approach, using the ideas of wavelet analysis, we keep wave vectors 
($k$) and scales ($a$) separately. We derive a renormalization group like 
equations, which can be used to study the energy distribution between 
different scales. The results can be applied for the investigation of 
turbulence velocity experimental data, as well for further theoretical 
research.

\centerline{***}
The author is grateful to Prof. G.A.Ososkov for critical reading the 
manuscript. I'm also grateful and to Profs. V.V.Ivanov and I.V.Puzynin 
for stimulating interest to this research. 

The work was supported in part by ESPRIT project 21042.  

\end{document}